\documentclass[letterpaper, 10 pt, conference]{ieeeconf}
\IEEEoverridecommandlockouts
\usepackage{cite}
\usepackage{amsmath,amssymb,amsfonts,mathtools}
\usepackage{algorithmic}
\usepackage{graphicx}
\usepackage{textcomp}
\usepackage[normalem]{ulem}

\usepackage[dvipsnames]{xcolor}

\let\labelindent\relax
\usepackage{enumitem}

\def\BibTeX{{\rm B\kern-.05em{\sc i\kern-.025em b}\kern-.08em
    T\kern-.1667em\lower.7ex\hbox{E}\kern-.125emX}}
\begin{document}
\newtheorem{lemma}{Lemma}
\newtheorem{theorem}{Theorem}
\newtheorem{definition}{Definition}
\newtheorem{assumption}{Assumption}
\newtheorem{corollary}{Corollary}
\newtheorem{remark}{Remark}
\newtheorem{algorithm}{Algorithm}
\newtheorem{conjecture}{Conjecture}

\title{\LARGE \bf
CVaR-based Safety Analysis in the Infinite Time Horizon Setting
}

\author{Chuanning Wei$^\dag$, Michael Fau{\ss}$^\ddag$, and Margaret P. Chapman$^\dag$ 
\thanks{$^\dag$C.W. and M.P.C. are with the Edward S. Rogers Sr. Department of Electrical and Computer Engineering, University of Toronto, Toronto, ON, Canada. Contact email: {\tt\small chuanning.wei@mail.utoronto.ca}, {\tt\small mchapman@ece.utoronto.ca}.}
\thanks{$^\ddag$M.F. is with the Department of Electrical and Computer Engineering, Princeton University, Princeton, NJ, USA. Contact email: {\tt\small mfauss@princeton.edu}.}
\thanks{C.W. was supported by a Natural Sciences and Engineering Research Council of Canada (NSERC) Undergraduate Student Research Award (USRA). This research was enabled in part by support provided by Compute Canada (www.computecanada.ca). M.P.C. and C.W. acknowledge support provided by the Edward S. Rogers Sr. Department of Electrical and Computer Engineering, University of Toronto.}
}

\maketitle
\pagestyle{empty}
\thispagestyle{empty}
\begin{abstract}
We develop a risk-averse safety analysis method for stochastic systems on discrete infinite time horizons. Our method quantifies the notion of risk for a control system in terms of the severity of a harmful random outcome in a fraction of the worst cases. In contrast, classical methods quantify risk in terms of the probability of a harmful event. Our theoretical arguments are based on the analysis of a value iteration algorithm on an augmented state space. We provide conditions to guarantee the existence of an optimal policy on this space. We illustrate the method numerically using an example from the domain of stormwater management.
\end{abstract}

\section{Introduction}\label{Introduction}
The standard approach to stochastic safety analysis is to minimize the probability that a control system violates a given safety or performance criterion. Variations of this problem have been studied in the context of non-adversarial disturbances \cite{abate2008probabilistic, summers2010verification}, adversarial disturbances \cite{ding2013stochastic}, distributional robustness \cite{yang2018dynamic}, and temporal logic \cite{sadigh2016safe, jha2018safe}. 

While minimizing the probability of a harmful event is useful, it may be imperative to quantify and minimize its severity directly. For example, in periods of heavy rainfall, stormwater overflows may be inevitable, but reducing the magnitude of the overflows (e.g., maximum flood level or overflow volume) is important to preserve the structural integrity of cities. For another example, adverse side effects from chemotherapy may be unavoidable, but when side effects are less severe, treatment protocols can continue more readily.
Also, the International Organization for Standardization (ISO) risk management guidelines include ``the likelihood of events and consequences'' and ``the nature and magnitude of consequences'' as factors for consideration in risk analysis \cite[Sec. 6.4.3]{isosource}. The importance of quantifying and minimizing the magnitude (i.e., severity) of a random harmful outcome has motivated the development of risk-averse safety analysis methods for control systems \cite{samuelson2018safety, chapmanACC, mpctacsubmission, chapman2021risk}.

A mathematical tool that accommodates both probability and magnitude is a \emph{risk functional}, which maps a random variable (representing a cost) to an extended real number. Early research on optimizing Markov decision processes (MDPs) with respect to a risk functional originated in 1972 and was formulated using the \emph{Exponential Utility} by Howard and Matheson \cite{howard1972risk}. Since then, various other risk functionals have been considered in the literature, including the \emph{Value-at-Risk} (VaR), \emph{Conditional Value-at-Risk} (CVaR), and \emph{Mean-Deviation} \cite{shapiro2021lectures}. Nevertheless, it is instructive to briefly describe the classical risk functional Exponential Utility.

The Exponential Utility functional is defined as 
\begin{equation}
    \rho_{\text{EU},\theta}(G) \coloneqq \textstyle{\frac{-2}{\theta}} \log E(\exp(\textstyle{\frac{-\theta}{2}}G)),
\end{equation}
where $G$ is a non-negative random variable, $\theta < 0$ in the risk-averse case, and $\theta > 0$ in the risk-seeking case. The Exponential Utility is not positively homogeneous. That is, $\lambda \rho_{\text{EU},\theta}(G)$ need not equal $\rho_{\text{EU},\theta}(\lambda G)$ for $\lambda > 0$. Under certain conditions, it holds that $\underset{\theta \rightarrow 0}{\lim} \rho_{\text{EU},\theta}(G) = E(G)$, and if $|\theta|$ is sufficiently small, then $\rho_{\text{EU},\theta}(G) \approx E(G) - \frac{\theta}{4}\text{variance}(G)$ \cite{whittle1981risk}. When this approximation is not valid, $\rho_{\text{EU},\theta}(G)$ may be difficult to interpret. Recently, we showed that using a more negative value for $\theta$ in the optimal control problem of minimizing $\rho_{\text{EU},\theta}(G)$ can yield a distribution of $G$ with a higher mean and a higher variance \cite{kevincontrolsystech}.

The VaR and CVaR have been considered as alternative, potentially more useful, risk functionals. The CVaR, in particular, is becoming popular in the control systems and robotics research communities due to its quantitative and intuitive interpretation \cite{van2015distributionally, samuelson2018safety, milleryang, majumdar2020should}. The CVaR quantifies the severity of a harmful outcome in a given fraction $\alpha$ of the worst cases. That is, if $G$ is a continuous random variable with finite $E(|G|)$, then the CVaR of $G$ at level $\alpha \in (0,1)$ is the expectation of $G$ in the $\alpha \cdot 100\%$ worst cases. As we will describe further in Section \ref{Prem}, the CVaR also satisfies the four desirable axioms proposed by Artzner et al. \cite{Artzner}. In contrast, the VaR is commonly criticized for lacking subadditivity, and Exponential Utility has the shortfalls we discussed above.



In prior work, we proposed a safety analysis framework that uses the CVaR functional to assess the magnitude of a random maximum cost incurred by a control system \cite{mpctacsubmission,chapman2021risk}. 
The theory from \cite{mpctacsubmission,chapman2021risk} concerns a discrete finite time horizon. In the current paper, we will study the infinite time horizon case. It will become clear that this extension is challenging and requires the development of some interesting theoretical arguments. 




The problem of risk-averse safety analysis for control systems is related to the problem of optimizing risk-averse MDPs. Both problems cannot necessarily be solved using dynamic programming (DP) recursions on the state space. This is because some risk functionals, including the CVaR, Mean-Variance, and Expected Utility (exception: Exponential Utility), do not satisfy an analogue of the law of iterated expectations, in which the current state summarizes sufficient information about the past.
A popular approach to mitigate this issue is to define the dynamics of an extra state so that a DP recursion or a linear program can be formulated on the augmented state space \cite{bauerle2011markov, bauerlerieder, bauerle2020minimizing, haskell, borkar2014risk, chapman2021risk}.
An optimal policy that depends on the augmented state dynamics can be constructed under a measurable selection condition \cite{bauerle2011markov, bauerlerieder, bauerle2020minimizing, chapman2021risk}; such a policy may be called an \emph{optimal precommitment policy} to highlight its extra dependencies.
One can avoid state-space augmentation in a CVaR setting when minimizing an expected cumulative cost subject to a CVaR constraint on a stage cost \cite{van2015distributionally, samuelson2018safety} or when minimizing the CVaR of a terminal cost \cite{milleryang}.

Here, we focus on optimizing an MDP in which the random cost is a \emph{supremum} of random stage costs over an infinite time horizon. In the MDP literature, it is more common to optimize a random \emph{cumulative} cost, e.g., see \cite{bauerle2011markov, bauerlerieder,bauerle2020minimizing}. A random cumulative cost represents a setting in which the severity of an undesirable outcome accumulates as a system evolves. However, there are cases when the severity of an undesirable outcome has an instantaneous nature, which motivates the use of a maximum cost over time. For example, in the application of stormwater management, a maximum water level indicates a maximum discharge rate, and this rate represents an instantaneous amount of stress on downstream infrastructure \cite{mpctacsubmission}. 
Moreover, the problem of optimizing a maximum cost incurred by the trajectory of a non-stochastic control system has been studied using Hamilton-Jacobi (HJ) reachability analysis over the past 15 years; e.g., see \cite{chen2018hamilton} and \cite{sylviathesis}. Historically, HJ analysis has been applied mainly to aerospace applications.

In this paper, we extend the risk-averse safety analysis method in \cite{chapman2021risk} to the infinite time horizon setting. The core problem is to minimize the CVaR of the supremum of stage costs subject to the dynamics of an MDP and construct an optimal precommitment policy under appropriate assumptions. The optimal values define a family of risk-averse safety specifications, which quantify the severity of a harmful outcome in a given fraction $\alpha$ of the worst cases. The extension 
necessitates some different techniques compared to \cite{chapman2021risk}. In particular, the solution requires:
\begin{enumerate}
    \item Deriving a forward DP recursion on an augmented state space (Lemma \ref{DPrecursion}) and
    \item Showing that a sequence of value functions converges pointwise to an optimal expected maximum cost (Theorem \ref{mainthm}).
\end{enumerate}

In addition to these technical contributions, we wish to highlight two conceptual implications of our work. First, in the finite time horizon case, the CVaR optimal policy is 
\emph{time-varying} \cite[Th. 2]{chapman2021risk}. 
However, in the infinite time horizon case, the 
optimal policy is \emph{time-invariant} and satisfies a time-independent, and therefore simpler, optimality equation (Theorem 1d). Once a time-invariant policy is available, it enjoys reduced memory requirements compared to a time-varying policy.
Second, a classical stochastic ``first hitting time'' reach-avoid problem is meaningful on an infinite time horizon \cite[Sec. 5.3]{summers2010verification}. In future work, we are interested in proposing and studying risk-averse reach-avoid problems on long time horizons. Such problems may be relevant for long-term planning of hydro-electric operations (e.g., generate enough electricity but also alleviate downstream flooding). The present work is a necessary stepping stone.

This paper is organized as follows. Section \ref{Prem} presents notation and background about CVaR. Section \ref{probstatement} states the problem of interest, and Section \ref{secIII} describes the problem-solving approach. Section \ref{theorysec} provides the theoretical results. Section \ref{numex} presents a numerical example, and Section \ref{conc} offers brief concluding remarks.

\section{Preliminaries}\label{Prem}
\subsection{Notation}
If $S$ is a metrizable space, then $\mathcal{B}_{S}$ is the Borel sigma algebra on $S$, 
$\mathcal{P}_{S}$ is the set of probability measures on $(S,\mathcal{B}_{S})$ with the weak topology, and $M^{+}_{S}$ is the set of non-negative Borel-measurable functions on $S$. If $y \in S$, then $\delta_y \in \mathcal{P}_{S}$ is the Dirac measure on $(S,\mathcal{B}_{S})$ concentrated at $y$. $\mathbb{N}$ is the set of natural numbers and $\mathbb{N}_0 \coloneqq \mathbb{N} \cup \{0 \}$. For $n \in \mathbb{N}$, $\mathbb{R}_+^n$ is the non-negative orthant in $\mathbb{R}^n$. $\mathbb{R}^* \coloneqq \mathbb{R} \cup \{-\infty,+\infty\}$ is the extended real line. For $p \in \mathbb{N} \cup\{+\infty\}$, $L^p(\Omega,\mathcal{F},\nu)$ is the $L^p$ space corresponding to the measure space $(\Omega,\mathcal{F},\nu)$. If $\nu$ is a probability measure, the notation $G \in L^p(\Omega,\mathcal{F},\nu)$ means that $G$ is a random variable defined on $(\Omega,\mathcal{F},\nu)$ whose $L^p$ norm is finite.
We use the following abbreviations: w.r.t.\ = with respect to, l.s.c. = lower semi-continuous, and a.e.\ = almost everywhere or almost every.

\subsection{Conditional Value-at-Risk}
Here, we present a standard definition for the CVaR \cite[Eq. (3.11)]{shapiro2012} and some of its important properties. 

\begin{definition}[Conditional Value-at-Risk]
Let a probability space $(\Omega,\mathcal{F},\mu)$ and a random variable $Y \in L^1(\Omega,\mathcal{F},\mu)$ be given; i.e., $Y$ is a random variable on $(\Omega,\mathcal{F},\mu)$ such that $E(|Y|) \coloneqq \int_{\Omega}|Y|\mathrm{d}\mu$ is finite. The \emph{Conditional
Value-at-Risk} of $Y$ at the risk-aversion level $\alpha \in (0,1]$ is defined by
\begin{equation}\label{defCVaR}
    \text{CVaR}_\alpha(Y) \coloneqq \underset{s \in \mathbb{R}}{\inf} \left( s + \textstyle{\frac{1}{\alpha}}E(\max\{Y - s, 0\})\right).
\end{equation}
\end{definition}

\vspace{1mm}The CVaR is related to the \emph{Value-at-Risk} (VaR), which is defined by
\begin{equation}
    \text{VaR}_\alpha(Y) \coloneqq \inf \{y \in \mathbb{R} : \mu(\{Y \leq y\})\geq 1 - \alpha\}
\end{equation}
for $\alpha \in (0,1)$. The set $\{Y \leq y\} \coloneqq \{\omega \in \Omega : Y(\omega) \leq y \}$ is a member of $\mathcal{F}$ because $Y$ is measurable w.r.t. $\mathcal{F}$ and $\mathcal{B}_{\mathbb{R}}$. The CVaR can be written as an integral of the VaR w.r.t. the risk-aversion level \cite[Th. 6.2]{shapiro2021lectures},
\begin{equation}\label{averagedef}
    \text{CVaR}_\alpha(Y) = \frac{1}{\alpha}  \int^{1}_{1-\alpha} \text{VaR}_{1-p}(Y) \; \mathrm{d}p, \quad \alpha \in (0,1),
\end{equation}
which explains why \emph{Average Value-at-Risk} is a synoymn for CVaR. If $\alpha \in (0,1)$ and the distribution function $F_{Y}(y) \coloneqq \mu(\{Y \leq y\})$ is continuous at the point $y = \text{VaR}_\alpha(Y)$, then the $\text{CVaR}_{\alpha}(Y)$ is the expectation of $Y$ conditioned on the event $\{Y \geq \text{VaR}_{\alpha}(Y)\}$ \cite[Th. 6.2]{shapiro2021lectures}: 
\begin{equation}\label{conditiondef}
    \text{CVaR}_\alpha(Y) = E(Y|Y \geq \text{VaR}_\alpha(Y)),
\end{equation}
which explains the name Conditional Value-at-Risk.
Equations \eqref{averagedef} and \eqref{conditiondef} provide 
expressions for the CVaR in terms of the VaR. In particular, Eq. \eqref{conditiondef} indicates that, under some assumptions, the $\text{CVaR}_\alpha(Y)$ quantifies the magnitude of $Y$ in the $\alpha \cdot 100\%$ of the worst cases, which appear in the upper tail of the distribution of $Y$.

The CVaR on $L^1(\Omega,\mathcal{F},\mu)$ for $\alpha \in (0,1]$ satisfies the four axioms that define the class of \emph{coherent} risk functionals, proposed by Artzner et al. \cite{Artzner}. For convenience, we list these properties below using the notation $\rho \coloneqq \text{CVaR}_\alpha$:
\begin{enumerate}
    \item \emph{Monotonicity}: if $Y_1(\omega) \leq Y_2(\omega)$ for almost every $\omega \in \Omega$ with respect to $\mu$, then $\rho(Y_1) \leq \rho(Y_2)$;
    \item \emph{Subadditivity}: $\rho(Y_1+Y_2) \leq \rho(Y_1) + \rho(Y_2)$;
    \item \emph{Translation Equivariance}: $\rho(Y+a) = \rho(Y) +a$ for all $a \in \mathbb{R}$;
    \item \emph{Positive Homogeneity}: $\rho(\lambda Y) = \lambda \rho(Y)$ for all $\lambda \geq 0$.
\end{enumerate}
A discussion about these axioms and why the VaR is not subadditive can be found in \cite[Sec. 2.2]{kisi}.

\section{Problem Statement}\label{probstatement}
We consider a stochastic control system operating on a discrete infinite time horizon $\mathbb{N}_0$. For all $t \in \mathbb{N}_0$, the realizations of the random state $X_t$, random control $U_t$, and random disturbance $W_t$ are elements of a non-empty Borel space, $S$, $C$, and $D$, respectively. The realizations of $X_0$ are concentrated at an arbitrary $x \in S$. The disturbance process $W_0,W_1,\dots$ is a sequence of random objects, such that for all $t \in \mathbb{N}_0$, given $(X_t,U_t)$, $W_t$ is independent of $W_\tau$ for all $\tau \neq t$. If $(x,u) \in S \times C$ is the realization of $(X_t,U_t)$, then the distribution of $W_t$ is $p(\cdot|x,u)$, where $p(\cdot|\cdot,\cdot)$ is a Borel-measurable stochastic kernel on $D$ given $S \times C$. In addition, the distribution of $X_{t+1}$ is $Q(\cdot|x,u)$, which is defined by
\begin{equation}
    Q(B|x,u) \coloneqq p(\{w \in D : f(x,u,w) \in B \} | x,u),
    \end{equation}
where $B \in \mathcal{B}_S$ and $f : S \times C \times D \rightarrow S$ is a Borel-measurable function.

We consider a random cost 
\begin{equation}\label{suprandcost}
    Y \coloneqq \underset{t \in \mathbb{N}_0}{\sup} \, c(X_t,U_t),
\end{equation}
where $c : S \times C \rightarrow \mathbb{R}$ is Borel measurable, bounded, and non-negative. In particular, we assume that $c(x,u) \in \mathcal{Z} \coloneqq [0,\bar{c}]$ for all $(x,u) \in S \times C$ with $\bar{c} > 0$. The problem is to compute a family of \emph{risk-averse safe sets}. A risk-averse safe set 
\begin{equation}\label{rssafeset}
    \mathcal{S}_\alpha^r \coloneqq \left\{ x \in S :  V_\alpha^*(x) \leq r \right\}, \quad \alpha \in (0,1], \quad r \in \mathcal{Z},
\end{equation}
is defined in terms of a CVaR-optimal control problem,\footnote{If $c$ is not non-negative, define $\tilde c \coloneqq c - \underline{b}$, where $\underline{b} \in \mathbb{R}$ is a lower bound for $c$, $\tilde Y \coloneqq \underset{t \in \mathbb{N}_0}{\sup} \, \tilde c (X_t,U_t) = Y - \underline{b}$, and $\tilde V_\alpha^*(x) \coloneqq \underset{\pi \in \Pi}{\inf} \text{CVaR}_{\alpha,x}^\pi(\tilde Y)$. One computes the criterion of interest $V_\alpha^*$ using the computation of $\tilde V_\alpha^*$ and the relation $ V_\alpha^* = \tilde V_\alpha^* + \underline{b}$, which holds as a result of translation equivariance.\label{footnote1}}
\begin{equation}\label{coreproblem}
    V_\alpha^*(x) \coloneqq \inf_{\pi \in \Pi} \text{CVaR}_{\alpha,x}^\pi(Y),
\end{equation}
where $Y$ is the supremum random cost defined by \eqref{suprandcost}.
Let us describe the other terms in \eqref{coreproblem}:
\begin{itemize}
    \item $\Pi$ is a class of history-dependent policies (to be defined in Section \ref{secIIIA}),
    \item $\alpha \in (0,1]$ is a risk-aversion parameter, and
    \item $\text{CVaR}_{\alpha,x}^\pi(Y)$ is the CVaR of $Y$ at level $\alpha$ when the system is initialized at $x$ and uses the policy $\pi$.
\end{itemize}

A risk-averse safe set $\mathcal{S}_\alpha^r$ \eqref{rssafeset} represents the set of initial states from which the expectation of $Y$ \eqref{suprandcost} in the $\alpha \cdot 100\%$ worst cases can be reduced to a threshold $r$. Hence, $\mathcal{S}_\alpha^r$ is a safety notion that permits \emph{flexibility} in the definition of ``the worst case'' and assesses the \emph{severity} of a random cost $Y$ for a stochastic control system. 
One may choose the stage cost $c$ to quantify a distance between a state realization and a desired operating region $K \in \mathcal{B}_S$. In this case, $Y$ represents a distance between the random state trajectory and $K$ in the long run. In our example of a stormwater system, we will define $c$ to quantify an overflow amount (Section \ref{numex}). 

\section{State-Space Augmentation Approach}\label{secIII}
\subsection{Defining a Control System on an Augmented State Space}\label{secIIIA}
While it is not possible to compute $V_\alpha^*$ \eqref{coreproblem} using a DP recursion on $S$, we will overcome this challenge by defining an augmented state $(X_t,Z_t)$, which has realizations in $S \times \mathcal{Z}$. The role of $Z_t$ is to record the running maximum up to time $t$. 
Formally, we define $X_t$, $U_t$, and $Z_t$ on the sample space
\begin{equation}
    \Omega \coloneqq (S \times \mathcal{Z} \times C)^\infty,
\end{equation}
where every $\omega \in \Omega$ takes the form 
\begin{equation}\label{myomega}
    \omega = (x_0,z_0,u_0,x_1,z_1,u_1,\dots),
\end{equation}
and the coordinates of $\omega$ are related causally. We define $X_t$, $Z_t$, and $U_t$ to be projections from $\Omega$ to $S$, $\mathcal{Z}$, and $C$, respectively, such that for all $\omega \in \Omega$ of the form in \eqref{myomega}, 
\begin{equation}
    X_t(\omega) \coloneqq x_t, \quad Z_t(\omega) \coloneqq z_t, \quad U_t(\omega) \coloneqq u_t.
\end{equation}
%
In addition, the dynamics of $Z_t$ are given by
\begin{equation}
    Z_{t+1} = \max\{ Z_{t}, c(X_t,U_t) \}, \quad t \in \mathbb{N}_0.
\end{equation}
The realizations of $Z_0$ are concentrated at a point $z \in \mathcal{Z}$. Later in our analysis, we will see that choosing $z =0$ is particularly useful. 

We define the random cost $Y : \Omega \rightarrow \mathbb{R}$ as follows: for all $\omega \in \Omega$ of the form in \eqref{myomega},
\begin{equation}\label{myformalY}
    Y(\omega) \coloneqq \underset{t \in \mathbb{N}_0}{\sup} \, c(X_t(\omega),U_t(\omega)) = \underset{t \in \mathbb{N}_0}{\sup} \, c(x_t,u_t).
\end{equation} 


$\Pi$ is the class of \emph{stationary} policies that are history-dependent through $(X_t,Z_t)$.
\begin{definition}[Policy class $\Pi$]
Any $\pi \in \Pi$ takes the form $\pi = (\mu,\mu,\dots)$, where $\mu(\cdot|\cdot,\cdot)$ is a Borel-measurable stochastic kernel on $C$ given $S \times \mathcal{Z}$.
\end{definition}

$\Pi'$ is the class of stationary and non-stationary policies that are history-dependent through $(X_t,Z_t)$.
\begin{definition}[Policy class $\Pi'$]
Any $\pi \in \Pi'$ takes the form $\pi = (\mu_0,\mu_1,\dots)$, where $\mu_t(\cdot|\cdot,\cdot)$ is a Borel-measurable stochastic kernel on $C$ given $S \times \mathcal{Z}$ for all $t \in \mathbb{N}_0$.
\end{definition}

In particular, $\Pi'$ is a superset of $\Pi$.

\begin{remark}[Evolution of the augmented system]\label{evremark}
Given a policy $\pi = (\mu_0,\mu_1,\dots) \in \Pi'$ and an initial augmented state $(x,z) \in S\times \mathcal{Z}$, the augmented system evolves as follows. Initialize $t = 0$ and $(x_0,z_0) = (x,z)$. For $t = 0,1,\dots$, repeat the following four steps:
\begin{enumerate}
    \item A realization $u_t$ of $U_t$ occurs according to $\mu_t(\cdot|x_t,z_t)$.
    \item A realization $w_t$ of $W_t$ occurs according to $p(\cdot|x_t,u_t)$.
    \item A realization $(x_{t+1},z_{t+1})$ of $(X_{t+1},Z_{t+1})$ is given by $(x_{t+1},z_{t+1}) = (f(x_t,u_t,w_t),\max\{z_t,c(x_t,u_t)\})$.
    \item Time $t$ updates by 1, and proceed to step 1.
\end{enumerate}
\end{remark}

Next, we present a family of probability measures on $(\Omega, \mathcal{B}_\Omega)$ that we use throughout the paper. 

\subsection{Probability Measures, $P_{x,z}^\pi$ and $P_{x,z}^{\pi,j}$}
Let $(x,z) \in S \times \mathcal{Z}$ and $\pi = (\mu_0,\mu_1,\dots) \in \Pi'$ be given. By \cite[Prop. 7.28]{bertsekas2004stochastic}, there is a unique probability measure $P_{x,z}^\pi \in \mathcal{P}_\Omega$, whose marginals satisfy useful properties. To describe the properties, the following notations are convenient:
\begin{align}
    \Omega_j & = (S \times \mathcal{Z} \times C)^j,\label{omegat}\\
    \bar{x}_i & = (x_i,z_i) \text{ or } \bar{x}_i = (z_i,x_i),\\
     \nu_{x,z}(\mathrm{d}\bar{x}_0) &  = \delta_{z}(\mathrm{d}z_0) \; \delta_x(\mathrm{d}x_0),
\end{align}
with $j \in \mathbb{N}$, $i \in \mathbb{N}_0$, $x_i \in S$, $x \in S$, $z_i \in \mathcal{Z}$, and $z \in \mathcal{Z}$. We denote the transition kernel on the augmented state space by
\begin{equation}\label{2b}\begin{aligned}
       \bar{Q}(\mathrm{d}\bar{x}_{i+1}|\bar{x}_{i},u_i) & = \delta_{\max\{z_i,c(x_i,u_i)\}}(\mathrm{d}z_{i+1}) \ Q(\mathrm{d}x_{i+1}|x_i,u_i).
\end{aligned}\end{equation}
For $j \in \mathbb{N}$, we denote the marginal of $P_{x,z}^\pi$ on $\Omega_j$ by $P_{x,z}^{\pi,j}$,
\begin{align}\label{Ptxz}
    P_{x,z}^{\pi,j}(A) & = P_{x,z}^\pi(H_{j}^{-1}(A)), \quad A \in \mathcal{B}_{\Omega_j},
\end{align}
where $H_j : \Omega \rightarrow \Omega_j$ is defined by
\begin{equation}\label{Ht}
    H_j  \coloneqq (X_0, Z_0, U_0,\dots,X_{j-1},Z_{j-1},U_{j-1}),
\end{equation}
and $H_{j}^{-1}(A)$ is defined by
\begin{equation}
    H_{j}^{-1}(A) \coloneqq \{H_j \in A \} \coloneqq \{\omega \in \Omega : H_{j}(\omega) \in A \}.
\end{equation}

Now, we are ready to state the property of $P_{x,z}^{\pi,j}$ \eqref{Ptxz} alluded above.
For all $j \in \mathbb{N}$, if $g : \Omega_j \rightarrow \mathbb{R}^*$ is Borel measurable and non-negative, then $\int_{\Omega_j} g \; \mathrm{d}P_{x,z}^{\pi,j}$ equals \eqref{marginal} \cite[Prop. 7.28]{bertsekas2004stochastic}; \eqref{marginal} is located at the top of the next page.
\begin{figure*}
    \begin{equation}
        \label{marginal}
     \int_{\Omega_j} \hspace{-1mm} g(\bar{x}_0,u_0,\dots,\bar{x}_{j-1},u_{j-1}) \  \mu_{j-1}(\mathrm{d}u_{j-1}|\bar{x}_{j-1}) \ \bar{Q}(\mathrm{d}\bar{x}_{j-1}|\bar{x}_{j-2},u_{j-2})  \cdots  \mu_{1}(\mathrm{d}u_{1}|\bar{x}_1) \ \bar{Q}(\mathrm{d}\bar{x}_1|\bar{x}_0,u_{0}) \ \mu_{0}(\mathrm{d}u_{0}|\bar{x}_0)\ \nu_{x,z}(\mathrm{d}\bar{x}_0)
    \end{equation}\vspace{-10mm}
\end{figure*}
\subsection{Evaluating Random Variables on $(\Omega,\mathcal{B}_\Omega,P_{x,z}^\pi)$}
If $G : \Omega \rightarrow \mathbb{R}^*$ is Borel measurable and non-negative, then the expectation of $G$ with respect to $P_{x,z}^\pi$ is defined by 
\begin{equation}
    E_{x,z}^\pi(G) \coloneqq \int_\Omega G \; \mathrm{d}P_{x,z}^\pi.
\end{equation}

The above definition is useful for defining the CVaR of $Y$ \eqref{myformalY} formally. Since $Y : \Omega \rightarrow \mathbb{R}$ is bounded everywhere and Borel measurable, we have that $Y \in L^1(\Omega,\mathcal{B}_\Omega, P_{x,z}^\pi)$ for all $(x,z) \in S\times \mathcal{Z}$ and $\pi \in \Pi'$. Considering $z = 0$, the CVaR of $Y \in L^1(\Omega,\mathcal{B}_\Omega, P_{x,0}^\pi)$ at level $\alpha \in (0,1]$ is given by 
\begin{equation}\label{cvarexpress}
    \text{CVaR}_{\alpha,x}^\pi(Y)  \coloneqq \inf_{s \in \mathbb{R}} \left( s + \textstyle \frac{1}{\alpha} E_{x,0}^\pi(\max\{Y - s,0\} ) \right). 
\end{equation}

\subsection{Outline of Theory}
We use \eqref{cvarexpress} to express $V_\alpha^*$ \eqref{coreproblem} as a bi-level optimization problem by exchanging the order of infima over $\Pi$ and $\mathbb{R}$. For all $s \in \mathbb{R}$ and $x \in S$, we define 
\begin{equation}\label{Vs}
    V_s(x) \coloneqq \inf_{\pi \in \Pi} E_{x,0}^\pi(\max\{Y - s,0\} ).
\end{equation}
Then, for all $\alpha \in (0,1]$ and $x \in S$, it holds that
\begin{equation}\label{valpha}
     V_\alpha^*(x) = \inf_{s \in \mathbb{R}} \left( s + {\textstyle \frac{1}{\alpha}} V_s(x)  \right) = \min_{s \in \mathcal{Z}} \left( s + {\textstyle \frac{1}{\alpha}} V_s(x)  \right).
\end{equation}
A minimizer $s_{\alpha,x}^* \in \mathcal{Z}$ exists due to $Y(\omega) \in \mathcal{Z}$ for all $\omega \in \Omega$, the continuity of $s \mapsto s + {\textstyle \frac{1}{\alpha}} V_s(x)$, and the compactness of $\mathcal{Z}$ \cite[Lemma 1]{chapman2021risk}. Next, we present a procedure to compute a family of risk-averse safe sets $\{\mathcal{S}_\alpha^r : \alpha \in \Lambda, r \in R\}$, where $\mathcal{S}_\alpha^r$ is defined by \eqref{rssafeset}, $\Lambda \subseteq (0,1]$, and $R \subseteq \mathcal{Z}$.
\begin{enumerate}
    \item For all $s \in \mathcal{Z}$, implement a value iteration algorithm on $S \times \mathcal{Z}$ to provide $V_s$ \eqref{Vs} in principle (Algorithm \ref{alg} and relevant theory to be presented in Section \ref{theorysec}).
    \item Use the family of functions $\{ V_s : s \in \mathcal{Z} \}$ to compute $ \{ V_\alpha^* : \alpha \in \Lambda \}$ by applying \eqref{valpha}.
    \item Use $ \{ V_\alpha^* : \alpha \in \Lambda\}$ to calculate $\{\mathcal{S}_\alpha^r : \alpha \in \Lambda, r \in R\}$.
\end{enumerate}

Our theory guarantees the computation of risk-averse safe sets and optimal precommitment policies exactly in principle under a measurable selection condition (Assumption \ref{assumption1}). We define a sequence of value functions $\{ v_t^s : t \in \mathbb{N}_0 \}$ parametrized by $s \in \mathbb{R}$ (Algorithm \ref{alg}). We show that each $v_t^s$ enjoys desirable properties and the limit 
\begin{equation}
    v^s(x,z) \coloneqq \underset{t \rightarrow \infty}{\lim} v_t^s(x,z)
\end{equation}
exists for all $(x,z) \in S \times \mathcal{Z}$ (Theorem \ref{analysisvaluefunc}). Then, in Theorem \ref{mainthm} and Corollary \ref{Vsps}, we show that $v^s = J_s$, where $J_s(x,0) = V_s(x)$ for all $x \in S$. 
Lastly, we guarantee the existence of a policy parametrized by $s$, from which we obtain an optimal precommitment policy parametrized by $x$ and $\alpha$ (Remark \ref{getoptimalpre}). 
%

\section{Theoretical Results}\label{theorysec}
We make the following assumption. 

\begin{assumption}[Measurable selection]\label{assumption1} It holds that
\begin{enumerate}
    \item The control space $C$ is compact.
    \item The dynamics function $f$ and the stage cost $c$ are continuous with $c(x,u) \in \mathcal{Z} \coloneqq [0,\bar{c}]$ $\forall (x,u) \in S \times C$.
    \item The disturbance kernel $p(\cdot|\cdot,\cdot)$ is a continuous stochastic kernel on $D$ given $S \times C$.
\end{enumerate}
\end{assumption}

A measurable selection condition is used to guarantee the existence of an optimal policy. To optimize an expected cumulative cost for an MDP, it is typical to assume that $C$ is compact, $f$ and $p$ are continuous, and $c$ is l.s.c. and bounded below \cite[Def. 8.7]{bertsekas2004stochastic}. For risk-aware MDPs, it is common to impose additional conditions on $c$. For instance, the works \cite{bauerlerieder} and \cite{haskell} both assume bounded positive costs. The cost-update operation in Algorithm \ref{alg} (to follow) is a composition rather than a summation.
Assuming that $c$ is continuous is a natural choice that helps preserve lower semi-continuity under the cost-update operation. 

\begin{algorithm}[Value iteration]\label{alg}
Let Assumption \ref{assumption1} hold, and let $s \in \mathbb{R}$ be given. For all $t \in \mathbb{N}_0$, define the functions $v_t^s$ on $S \times \mathcal{Z}$ as follows: $v_0^s(x,z) \coloneqq \max\{z-s,0\}$ and 
\begin{multline*}
    v_{t+1}^s(x,z) \coloneqq \\ 
    \inf_{u \in C} \, \int_D v_{t}^s(f(x,u,w), \max\{z,c(x,u)\}) \, p(\mathrm{d}w|x,u). 
\end{multline*}
\end{algorithm}

 \vspace{3mm} We will exemplify a stopping criterion in Section \ref{numex}. The next definition is useful for analyzing Algorithm \ref{alg}. 
\begin{definition}[Operator $\Phi$, Borel space $F$]\label{Phioperator}
Define $F \coloneqq  S \times \mathcal{Z} \times C$. If $v \in  M_{S \times \mathcal{Z}}^+$, then we define $\Phi(v)$ by
\begin{equation*}
    \Phi(v)(x,z,u) \coloneqq \! \int_D \!\! v( f(x,u,w), \max\{z, c(x,u) \}) \ p(\mathrm{d}w|x,u)
\end{equation*}
for all $(x,z,u) \in F$. 
\end{definition}


Our first result guarantees regularity properties of the value functions of Algorithm \ref{alg}. 
\begin{theorem}[Analysis of Algorithm \ref{alg}]\label{analysisvaluefunc}
Let Assumption \ref{assumption1} hold, $s \in \mathbb{R}$, $\{v_t^s : t \in \mathbb{N}_0\}$ be given by Algorithm \ref{alg}, and $\bar{c}^s \coloneqq \max\{\bar{c}-s,0\}$. Then, the following statements hold: 
\begin{itemize}
    \item[\textbf{a)}] For all $t \in \mathbb{N}_0$, $v_t^s$ is l.s.c. and $0 \leq v_t^s \leq \bar{c}^s$;
    \item[\textbf{b)}] For all $(x,z) \in S \times \mathcal{Z}$, the limit of $\{v_t^s(x,z)\}_{t = 0}^\infty$ exists, which we denote by $v^s(x,z) \coloneqq \underset{t \rightarrow \infty}{\lim} v_t^s(x,z)$, and $0 \leq v_t^s \leq v_{t+1}^s \leq v^s \leq \bar{c}^s$ for all $t \in \mathbb{N}_0$; 
    \item[\textbf{c)}] $v^s$ is l.s.c. and $v^s(x,z) = \underset{u \in C}{\inf} \, \Phi(v^s)(x,z,u)$ for all $(x,z) \in S \times \mathcal{Z}$; 
    \item[\textbf{d)}] There is a Borel-measurable function $\kappa^s : S \times \mathcal{Z} \rightarrow C$ s.t. $v^s(x,z) = \Phi(v^s)(x,z,\kappa^s(x,z))$ $\forall (x,z) \in S \times \mathcal{Z}$. 
\end{itemize}
\end{theorem}

\hspace{-7mm}\begin{proof}
Part \textbf{a)} follows by induction, applying the arguments from \cite[Lemma 4]{chapman2021risk} and the fact that the infimum of an l.s.c. function over a compact metrizable space is l.s.c. \cite[Prop. 7.32 (a)]{bertsekas2004stochastic}. 
Also, for all $(x,z) \in S \times \mathcal{Z}$, there is a minimizer that attains $\underset{u \in C}{\inf} \, \Phi(v^s_{t})(x,z,u)$, and hence, $v^s_{t+1}(x,z) = \underset{u \in C}{\min} \, \Phi(v^s_{t})(x,z,u)$ \cite[Prop. 7.32 (a)]{bertsekas2004stochastic}. 

For part \textbf{b)}, the statement $v_t^s \leq v_{t+1}^s$ for all $t \in \mathbb{N}_0$ holds by induction, which we omit due to limited space. In particular, note that $v_1^s(x,z) = \underset{u \in C}{\inf} \, \max\{\max\{c(x,u),z\}-s,0\}$. The monotonicity and boundedness of the sequence of functions guarantees the existence of the limit. 
%
%

For part \textbf{c)}, $v^s$ is l.s.c. because it is a supremum of a family of l.s.c. functions. Since $\Phi(v^s)$ and $\Phi(v^s_t)$ are l.s.c. for all $t \in \mathbb{N}_0$ and $C$ is compact, $\Phi(v^s)$ and $\Phi(v^s_t)$ are inf-compact.\footnote{A function $\varphi : S \times \mathcal{Z} \times C \rightarrow \mathbb{R}$ is \emph{inf-compact}, if for all $(x,z) \in S \times \mathcal{Z}$ and $r \in \mathbb{R}$, the set $\{u \in C : \varphi(x,z,u) \leq r \}$ is compact.} A key step is that $\Phi(v^s) : S \times \mathcal{Z} \times C \rightarrow \mathbb{R}$ being l.s.c. implies that $\{u \in C : \Phi(v^s)(x,z,u) \leq r\}$ is closed for all $(x,z) \in S \times \mathcal{Z}$ and $r \in \mathbb{R}$. Hence, the limit as $t \rightarrow \infty$ and the minimum over $C$ commute \cite[Lemma 4.2.4]{hernandez2012discrete}.  
Now, $v^s(x,z) 
     = \underset{u\in C}{\min} \, \underset{t \rightarrow \infty}{\lim}  \int_D v_{t-1}^s(f(x,u,w), \max\{z,c(x,u)\}) p(\mathrm{d}w|x,u)$,
and for any $(x,z,u) \in F$, 
\begin{multline}
    v_{t-1}^s(f(x,u,\cdot),\max\{z,c(x,u)\}) \uparrow \\ v^s(f(x,u,\cdot),\max\{z,c(x,u)\}).
\end{multline}
Then, the desired result follows from the Monotone Convergence Theorem.

Part \textbf{d)} holds by using $v^s(x,z) = \underset{u \in C}{\inf} \, \Phi(v^s)(x,z,u)$ for all $(x,z) \in S \times \mathcal{Z}$, lower semi-continuity of $\Phi(v^s)$, and compactness of $C$ with \cite[Prop. 7.33]{bertsekas2004stochastic}. 
\end{proof}

To continue studying $v^s$, some additional information is needed. We define the random variable $Z : \Omega \rightarrow \mathbb{R}$ by
\begin{equation}
    Z \coloneqq \sup\{Z_0,c(X_0,U_0),c(X_1,U_1),\dots\},
\end{equation}
and we let $(x,z) \in S \times \mathcal{Z}$, $\pi \in \Pi'$, and $s \in \mathbb{R}$ be given. $Z$ and $\max\{Z_t-s,0\}$ for all $t \in \mathbb{N}_0$ are elements of $L^\infty(\Omega, \mathcal{B}_\Omega, P_{x,z}^\pi)$. 
This is because $c : S \times C \rightarrow \mathbb{R}$ is bounded, $X_t : \Omega \rightarrow S$, $Z_t : \Omega \rightarrow \mathcal{Z}$, $U_t : \Omega \rightarrow C$, and $c$ are Borel measurable, $\mathcal{Z} \subseteq \mathbb{R}$ is bounded, and $Z$ is the pointwise supremum of countably many functions. 
Also, we define 
\begin{align}
    J_{t,s}^\pi(x,z) & \coloneqq E_{x,z}^\pi(\max\{Z_t - s,0\}), \quad t \in \mathbb{N}_0, \label{Jtspi}\\
    J_s^\pi(x,z) & \coloneqq E_{x,z}^\pi(\max\{Z - s,0\}), \label{Jspi}\\
    J_s(x,z) & \coloneqq \inf_{\pi \in \Pi} J_s^\pi(x,z). \label{Js}
\end{align}
$J_{t,s}^\pi$ is a Borel-measurable function on $S \times \mathcal{Z}$, e.g., use \cite[Th. 4.1.11]{dudley2018real} and \cite[Prop. 7.29]{bertsekas2004stochastic}, and it holds that $0 \leq J_{t,s}^\pi \leq \bar{c}^s$. 
If $z = 0$, then $Y = Z$ a.e. w.r.t. $P_{x,0}^\pi$ due to the stage cost $c$ being non-negative and the realizations of $Z_0$ being concentrated at 0.
%
Hence, we have
\begin{equation}\label{my9}
    J_s^\pi(x,0) = E_{x,0}^\pi(\max\{Z - s,0\}) = E_{x,0}^\pi(\max\{Y - s,0\}),
\end{equation}
which is useful for a later result (Corollary \ref{Vsps}).
The next theorem specifies the relationship between $v^s$ and $J_s$. 

\begin{theorem}[$v^s = J_s$]\label{mainthm}
Let Assumption \ref{assumption1} hold. For all $s \in \mathbb{R}$, we have that $v^s = J_s$. 
\end{theorem}

To prove Theorem \ref{mainthm}, we require some preliminaries.
\begin{definition}[DP operator]\label{DPoperator}
Given $\mu$, a Borel-measurable stochastic kernel on $C$ given $S \times \mathcal{Z}$, 
the operator $T_{\mu} : M^{+}_{S \times \mathcal{Z}} \rightarrow M^{+}_{S \times \mathcal{Z}}$ is defined in \eqref{eq:T_Operator} on the next page.
\begin{figure*}
    \begin{equation}
        \label{eq:T_Operator}
        T_{\mu}(v)(x,z)  \coloneqq \int_C \int_S \int_{\mathcal{Z}} v(y,q) \ \delta_{\max\{z,c(x,u)\}}(\mathrm{d}q) \ Q(\mathrm{d}y|x,u) \ \mu(\mathrm{d}u|x,z)
    \end{equation}\vspace{-10mm}
\end{figure*}
Note that the variable $(x,z)$ in $T_{\mu}(v)(x,z)$ corresponds to the outer-most measure $\mu(\cdot|x,z)$.
\end{definition}

The next lemma provides a forward dynamic programming recursion on $S \times \mathcal{Z}$. Its proof is in the Appendix.
\begin{lemma}[DP recursion]\label{DPrecursion}
Let $\pi = (\mu_0,\mu_1,\dots) \in \Pi'$, $s \in \mathbb{R}$, and $t \in \mathbb{N}_0$ be given. 
It holds that $J_{0,s}^\pi = v_0^s$. If $t \geq 1$, then $J_{t,s}^\pi(x,z) = T_{\mu_0}( T_{\mu_1} ( \cdots ( T_{\mu_{t-1}}(v_0^s) ) \cdots ))(x,z)$ for all $(x,z) \in S \times \mathcal{Z}$. If $\pi = (\mu, \mu, \dots) \in \Pi$, then $J_{t+1,s}^\pi(x,z) = T_{\mu}(J_{t,s}^\pi)(x,z)$ for all $(x,z) \in S \times \mathcal{Z}$ and $t \in \mathbb{N}_0$.
\end{lemma}


The next definition uses the existence of a Borel-measurable selector $\kappa^s : S \times \mathcal{Z} \rightarrow C$ from Theorem \ref{analysisvaluefunc}.
\begin{definition}[$\pi^s$]\label{ps}
We define $\pi^s \coloneqq (\delta_{\kappa^s},\delta_{\kappa^s},\dots) \in \Pi$, which is a deterministic stationary policy. That is, if $(x_t,z_t) \in S \times \mathcal{Z}$ is the realization of $(X_t,Z_t)$, then the realizations of $U_t$ are concentrated at $\kappa^s(x_t,z_t) \in C$. 
\end{definition}

\begin{lemma}[$J_{t,s}^{\pi^s} \leq v^s$]\label{pslemma}
Let Assumption \ref{assumption1} hold, and let $s \in \mathbb{R}$ be given. 
Then, $J_{t,s}^{\pi^s} \leq v^s$ for all $t \in \mathbb{N}_0$. 
\end{lemma}
\hspace{-3.5mm}\begin{proof}
Proceed by induction. 
The base case is $J_{0,s}^{\pi^s} = v_0^s \leq v^s$ by Lemma \ref{DPrecursion} and Theorem \ref{analysisvaluefunc}.
Now, assume (the induction hypothesis) that for some $t \in \mathbb{N}_0$, it holds that $J_{t,s}^{\pi^s} \leq v^s$. Let $(x,z) \in S \times \mathcal{Z}$ be given. 
By Theorem \ref{analysisvaluefunc}, 
$v^s(x,z) = \Phi(v^s)(x,z,\kappa^s(x,z)) = T_{\delta_{\kappa^s}}(v^s)(x,z)$.
Since $0 \leq J_{t,s}^{\pi^s} \leq v^s$ and $J_{t,s}^{\pi^s}$ and $v^s$ are Borel-measurable functions on $S \times \mathcal{Z}$, it follows that 
$v^s(x,z) \geq T_{\delta_{\kappa^s}}(J_{t,s}^{\pi^s})(x,z) = J_{t+1,s}^{\pi^s}( x, z)$, where the equality holds by Lemma \ref{DPrecursion}, as $\pi^s$ is stationary.
\end{proof}

\begin{lemma}[$v_t^s \leq J_{t,s}^\pi$]\label{part2thm2}
Let Assumption \ref{assumption1} hold, and let $s \in \mathbb{R}$ be given. 
Then, $v_t^s \leq J_{t,s}^\pi$ for all $t \in \mathbb{N}_0$ and $\pi \in \Pi$.
\end{lemma}
\hspace{-3.5mm}\begin{proof}
Proceed by induction. The base case is $v_0^s = J_{0,s}^\pi$ for all $\pi \in \Pi$ by Lemma \ref{DPrecursion}. Now, assume (the ind. hyp.) that for some $t \in \mathbb{N}_0$, it holds that $v_t^s \leq J_{t,s}^\pi$ for all $\pi \in \Pi$. Let $(x,z) \in S \times \mathcal{Z}$ and $\pi = (\mu, \mu, \dots) \in \Pi$ be given. We have $J_{t+1,s}^\pi(x,z) = T_{\mu}(J_{t,s}^\pi)(x,z)$ from Lemma \ref{DPrecursion}, and since $0 \leq v_t^s \leq J_{t,s}^\pi$ by Theorem \ref{analysisvaluefunc} and the induction hypothesis, it follows that $J_{t+1,s}^\pi(x,z) \geq T_{\mu}(v_t^s)(x,z)$. Next, we apply the definitions of $v_{t+1}^s$ and $Q$ to find that for all $u \in C$, 
   $ v_{t+1}^s(x,z) \leq \textstyle \int_S \int_{\mathcal{Z}} v_{t}^s(y,q) \delta_{\max\{z,c(x,u)\}}(\mathrm{d}q) Q(\mathrm{d}y|x,u)$.
By integrating over all $u \in C$ with respect to $\mu(\cdot|x,z)$, we obtain that $v_{t+1}^s(x,z) \leq T_{\mu}(v_{t}^s)(x,z)$. All together, we conclude that
  $  v_{t+1}^s(x,z) \leq T_{\mu}(v_{t}^s)(x,z) \leq J_{t+1,s}^\pi(x,z)$.
\end{proof}

The proof of Theorem \ref{mainthm} follows.

\hspace{-7mm}\begin{proof}[Theorem \ref{mainthm}]
First, we show that $J_s \leq v^s$. 
For any $\pi \in \Pi'$ and $(x,z) \in S \times \mathcal{Z}$, it holds that $\underset{t \rightarrow \infty}{\lim} J_{t,s}^\pi(x,z) = J_s^\pi(x,z)$. Indeed, $0 \leq \max\{Z_t - s,0\} \uparrow \max\{Z - s,0\}$, which implies that $E_{x,z}^\pi(\max\{Z_t - s,0\}) \uparrow E_{x,z}^\pi(\max\{Z - s,0\})$ by the Monotone Convergence Theorem, where we use the measure space $(\Omega,\mathcal{B}_\Omega, P_{x,z}^\pi)$. By Lemma \ref{pslemma}, there is a policy $ \pi^s \in \Pi \subseteq \Pi'$ such that $J_{t,s}^{\pi^s} \leq v^s$ $\forall t \in \mathbb{N}_0$. Thus, $J_s^{\pi^s} = \underset{t \rightarrow \infty}{\lim} J_{t,s}^{\pi^s} \leq v^s$. Since $J_s \leq J_s^{\pi^s}$, we conclude that $J_s \leq v^s$.

Second, we show that $v^s \leq J_s$. A sufficient condition is $v_t^s \leq J_{t,s}^\pi$ $\forall t \in \mathbb{N}_0$ $\forall \pi \in \Pi$, which holds by Lemma \ref{part2thm2}. Indeed, this statement implies that $\underset{t \rightarrow \infty}{\lim} v_t^s \leq \underset{t \rightarrow \infty}{\lim} J_{t,s}^\pi$ $\forall \pi \in \Pi$, where the limits exist by previous analyses. Since $\underset{t \rightarrow \infty}{\lim} v_t^s = v^s$ and $\underset{t \rightarrow \infty}{\lim} J_{t,s}^\pi = J_{s}^\pi$ $\forall \pi \in \Pi'$, we have $v^s \leq J_{s}^\pi$ $\forall \pi \in \Pi$. Then, the desired result follows, $v^s \leq J_{s} = \underset{\pi \in \Pi}{\inf} J_{s}^\pi$.
%
%
\end{proof}

The last result explains how to obtain $V_s$ \eqref{Vs} using $J_s = v^s = \underset{t \rightarrow \infty}{\lim} v_t^s$ and provides an interpretation for $\pi^s$.
\begin{corollary}[Computing $V_s$, interpreting $\pi^s$]\label{Vsps}
Let Assumption \ref{assumption1} hold, $s \in \mathbb{R}$ be given, and $\pi^s$ satisfy Definition \ref{ps}. 
Then, $J_s(x,z) = E_{x,z}^{\pi^s}(\max\{Z - s,0\} )$ for all $ (x,z) \in S\times \mathcal{Z}$.
In particular, $V_s(x) = J_s(x,0) = E_{x,0}^{\pi^s}(\max\{Y - s,0\} )$ for all $ x \in S$.
\end{corollary}
\hspace{-3.5mm}\begin{proof}
Recall from the proof of Theorem \ref{mainthm} that $J_s \leq J_s^{\pi^s} = \underset{t \rightarrow \infty}{\lim} J_{t,s}^{\pi^s} \leq v^s$ (first part) and $v^s \leq J_s$ (second part). Therefore, $J_s = J_s^{\pi^s}$. More explicitly, we use \eqref{Jspi} to write $J_s(x,z) = E_{x,z}^{\pi^s}(\max\{Z - s,0\})$ for all $ (x,z) \in S \times \mathcal{Z}$.
Now, let $x \in S$ be given. By substituting $z = 0$, we obtain $J_s(x,0) = E_{x,0}^{\pi^s}(\max\{Z - s,0\} )$. Since \eqref{my9} holds for any $\pi \in \Pi'$ 
and $\pi^s$ is an element of $\Pi'$, it follows that
\begin{equation}\label{my8}
    J_s(x,0) = E_{x,0}^{\pi^s}(\max\{Z - s,0\} ) \overset{\eqref{my9}}{=} E_{x,0}^{\pi^s}(\max\{ Y - s,0\}).
\end{equation}
In addition, we have 
\begin{equation}\label{my11}\begin{aligned}
    J_s(x,0) &  \overset{\eqref{Js}}{=} \inf_{\pi \in \Pi} E_{x,0}^\pi(\max\{Z - s,0\}) \\
     & \overset{\eqref{my9}}{=}  \inf_{\pi \in \Pi} E_{x,0}^\pi(\max\{ Y - s,0\}) 
      \overset{\eqref{Vs}}{=} V_s(x).
\end{aligned}\end{equation}
Hence, $ V_s(x) \overset{\eqref{my11}}{=} J_s(x,0) \overset{\eqref{my8}}{=} E_{x,0}^{\pi^s}(\max\{ Y - s,0\})$.
\end{proof}

Corollary \ref{Vsps} indicates that under Assumption \ref{assumption1}, for each $s \in \mathbb{R}$, there is a deterministic policy $\pi^s \in \Pi$ that attains the infimum $V_s(x)$ for all $x \in S$. 
Next, we explain how to find an optimal policy that is parametrized by $\alpha$ and $x$.

\begin{remark}[Policy synthesis]\label{getoptimalpre}
Let $\alpha \in (0,1]$ and $x \in S$ be given. Recall that there is a minimizer $s_{\alpha,x}^* \in \mathcal{Z}$ such that the minimum CVaR is $V_\alpha^*(x) \overset{\eqref{valpha}}{=} \underset{s \in \mathbb{R}}{\inf} \left( s + {\textstyle \frac{1}{\alpha}} V_s(x)  \right) = s_{\alpha,x}^* + {\textstyle \frac{1}{\alpha}} V_{s_{\alpha,x}^*}(x)$. By Corollary \ref{Vsps}, for all $s \in \mathbb{R}$, we have $V_{s}(x) = E_{x,0}^{\pi^{s}}(\max\{ Y - s,0\})$, where $\pi^s = (\delta_{\kappa^s},\delta_{\kappa^s},\dots) \in \Pi$ satisfies Definition \ref{ps}. Select $s = s_{\alpha,x}^*$ to obtain an optimal precommitment policy $\pi^{s_{\alpha,x}^*} \in \Pi$. To deploy this policy, follow the procedure provided in Remark \ref{evremark}; use the initialization $(x_0,z_0) = (x,0)$, and for $t \in \mathbb{N}_0$, use the control $u_t = \kappa^{s_{\alpha,x}^*}(x_t,z_t)$. 
\end{remark}

\section{Numerical Example}\label{numex}
\begin{figure*}[!htb]
        \centering
        
        \includegraphics[width=\textwidth]{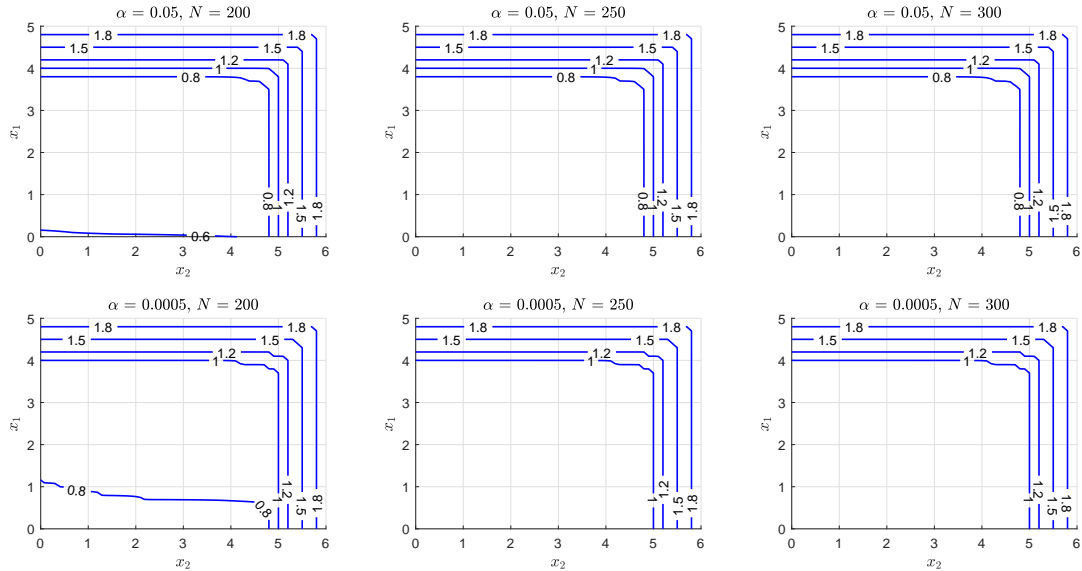}
        \caption{
        Contours of computations of risk-averse safe sets $\mathcal{S}_\alpha^r$ for $\alpha = 0.05$ (top row), $\alpha = 0.0005$ (bottom row), and $r \in \{0.2,0.5,0.6,0.8,1,1.2,1.5,1.8\}$. The results were computed using, from left to right, $N = 200$, $N=250$, and $N=300$ iterations. For sufficiently large $r$, the safe sets converge to approximately rectangular regions, in which the initial water levels are low enough so the system can operate within the given CVaR specifications indefinitely. For smaller $r$, however, safe operation becomes impossible on an infinite time horizon.}
        \label{fig1}
      \end{figure*}

We consider an urban stormwater system, consisting of two tanks connected by an automated valve, which we have adopted from our prior work \cite{chapman2021risk}. Water enters the system due to a random process of surface runoff, and water discharges through a storm sewer drain in tank 2 or through outlets that lead to a combined sewer. We penalize the latter discharge through a state-dependent stage cost $c(x,u) \coloneqq \max\{x_1 - k_1, x_2 - k_2, 0 \}$. The $i$th coordinate $x_i$ of the state $x = [x_1,x_2]^T$ is the water level of tank $i$, $k_i$ is the maximum water level prior to release into a combined sewer outlet ($k_1 = 3$ ft, $k_2 = 4$ ft), and the control $u \in C \coloneqq [0,1]$ is the valve position. We have implemented Algorithm \ref{alg} by discretizing the state space $S = [0, 5] \times [0, 6]$ ft\textsuperscript{2} at a resolution of $\Delta x = \frac{1}{10}$ ft in each dimension to estimate $\{v_0^s,v_1^s,\dots,v_{N}^s\}$ with $N \in \mathbb{N}$.\footnote{Our code is in MATLAB (The Mathworks, Inc.) and is available from https://github.com/mifauss/RSSAVSA-Infinite-Horizon/; the repository from \cite{chapman2021risk} provided its foundation. We have used a grid with $51\times 61 \times 21 = 65,331$ nodes to approximate $S \times \mathcal{Z}$, where $S = [0, 5] \times [0, 6]$ ft\textsuperscript{2} and $\mathcal{Z} = [0, 2]$ ft. We have used 21 values for $s$, $s \in \{0, 0.1, \dots, 2\}$, and a discrete distribution for the random surface runoff (mean: 2 cfs, variance: 0.3 cfs\textsuperscript{2}); cfs means cubic feet per second. We report approximate resources for an unoptimized implementation on Compute Canada's Cedar cluster with $N = 300$ iterations (number of cores: $32$, runtime: $81$ hours).} Let $\hat{V}_{\alpha,N}^*$ denote an estimate for $V_{\alpha}^*$ using $v_{N}^s$, and define $\gamma_\alpha(N',N) \coloneqq \sup\{ |\hat{V}_{\alpha,{N'}}^*(x) -\hat{V}_{\alpha,{N}}^*(x)| : x \in S\}$ for $N' <N$. We consider $N$ to be sufficiently large when our estimate for  $\gamma_\alpha(N',N)$ is at most $\frac{\Delta x}{20} = 0.005$ ft for some $N'<N$, and this serves as our stopping criterion for Algorithm \ref{alg}. Estimates for risk-averse safe sets $\mathcal{S}_\alpha^r$ for $\alpha = 0.05$ and $\alpha = 0.0005$ are shown in Figure \ref{fig1}, and sets prior to convergence are shown for comparison.
In particular, when $N'=280$ and $N=300$, we find that $\gamma_{0.05}(N',N) = 0.0030$ and $\gamma_{0.0005}(N',N) = 0.0048$. 
%


%

\section{Conclusions}\label{conc}
We have developed a CVaR-based safety analysis method for infinite time stochastic systems with theoretical guarantees. In the future, we plan to investigate the feasibility of grid-free policy improvement methods, e.g., stochastic rollout, to improve the scalability to high-dimensional systems, such as city-wide water networks.

\section*{Appendix}
Below, we provide a proof for Lemma \ref{DPrecursion}.\\
\indent\hspace{-7mm}\begin{proof}
Let $\pi = (\mu_0,\mu_1,\dots) \in \Pi'$, $s \in \mathbb{R}$, and $(x,z) \in S \times \mathcal{Z}$ be given. Recall the relations from the main text:
\begin{align*}
    \Omega_j & \overset{\eqref{omegat}}{=} (S \times \mathcal{Z} \times C)^j, \quad j \in \mathbb{N}, \\
    P_{x,z}^{\pi,j}(A) & \overset{\eqref{Ptxz}}{=} P_{x,z}^\pi(H_{j}^{-1}(A)), \quad A \in \mathcal{B}_{\Omega_j}, \quad j \in \mathbb{N},\\
       H_j & \overset{\eqref{Ht}}{=} (X_0, Z_0, U_0,\dots,X_{j-1},Z_{j-1},U_{j-1}), \quad j \in \mathbb{N},\\
        J_{t,s}^\pi(x,z) & \overset{\eqref{Jtspi}}{=} \int_\Omega \max\{Z_t - s,0\} \ \mathrm{d}P_{x,z}^\pi, \quad t \in \mathbb{N}_0.
\end{align*}
Let $t \in \mathbb{N}_0$ be given. Define $\varphi_{t,s} : \Omega_{t+1} \rightarrow \mathbb{R}$ by
\begin{equation}
    \varphi_{t,s}(x_0,z_0,u_0,\dots,x_t,z_t,u_t) \coloneqq \max\{z_t - s, 0\}.
\end{equation}
Note that $\varphi_{t,s}$ and $H_{t+1}$ are Borel measurable, $\varphi_{t,s}$ is non-negative, and $\varphi_{t,s} \circ H_{t+1} = \max\{Z_t - s, 0\}$. Thus,
\begin{equation}\label{myapp11}
    J_{t,s}^\pi(x,z) =  \int_\Omega \varphi_{t,s} \circ H_{t+1} \; \mathrm{d}P_{x,z}^\pi = \int_{\Omega_{t+1}} \varphi_{t,s} \; \mathrm{d}P_{x,z}^{\pi,t+1},
\end{equation}
where the second equality in \eqref{myapp11} holds by \cite[Th. 4.1.11]{dudley2018real}, as $P_{x,z}^{\pi,t+1}$ \eqref{Ptxz} is an image measure of $P_{x,z}^\pi$. 
We recall that for all $j \in \mathbb{N}$, if $g : \Omega_j \rightarrow \mathbb{R}^*$ is Borel measurable and non-negative, then $\int_{\Omega_j} g \; \mathrm{d}P_{x,z}^{\pi,j}$ is given by \eqref{marginal}.
If $t = 0$, then we apply the above statement and \eqref{myapp11} to find that
\begin{align}
    J_{0,s}^\pi(x,z) \hspace{-.5mm} & = \hspace{-1mm}\int_{\Omega_1} \hspace{-2mm}\max\{z_0 - s, 0\}  \mu_0(\mathrm{d}u_0|x_0,z_0)  \delta_z(\mathrm{d}z_0)  \delta_x(\mathrm{d}x_0) \nonumber\\
    & =  \max\{z - s, 0\}\\
    & = v_0^s(x,z).
\end{align}
The equality $J_{1,s}^\pi = T_{\mu_0}(v_0^s)$ follows from similar steps as described below. 
%
Let $t \in \{2,3,\dots\}$ be given, and use \eqref{marginal} and \eqref{myapp11} to find that
\begin{equation}\label{myfirst}\begin{aligned}
     J_{t,s}^\pi(x,z) 
   & = \textstyle \int_{S \times \mathcal{Z}}\int_{C} \int_{S\times \mathcal{Z}} \cdots \int_{S \times \mathcal{Z}}\int_C \; \max\{z_t - s, 0\}\\ & \hphantom{==}  \mu_{t}(\mathrm{d}u_{t}|\bar{x}_{t})\; \bar{Q}(\mathrm{d}\bar{x}_{t}|\bar{x}_{t-1},u_{t-1}) \cdots \bar{Q}(\mathrm{d}\bar{x}_1|\bar{x}_0,u_{0}) \\ & \hphantom{==}  \mu_{0}(\mathrm{d}u_{0}|\bar{x}_0) \; \nu_{x,z}(\mathrm{d}\bar{x}_0),
\end{aligned}\end{equation}
where there are $t+1$ copies of $S \times \mathcal{Z} \times C$ in \eqref{myfirst}. Since $\max\{z_t - s, 0\} = v_0^s(x_t,z_t)$ is constant in $(x_t,u_t)$ and from the definition of $\bar{Q}$ and $\nu_{x,z}$, it follows that
\begin{equation}\label{my1111}\begin{aligned}
    & J_{t,s}^\pi(x,z) = \\ &  \textstyle \int_{C} \int_{S} \int_ {\mathcal{Z}} \cdots\hspace{-.5mm} \Big(\hspace{-.5mm}  \int_{C} \int_{S} \int_{\mathcal{Z}}v_0^s(x_t,z_t) \;
   \delta_{\max\{z_{t-1},c(x_{t-1},u_{t-1})\}}(\mathrm{d}z_t)\\ &  \hphantom{=}  Q(\mathrm{d}x_t|x_{t-1},u_{t-1}) \; \mu_{t-1}(\mathrm{d}u_{t-1}|x_{t-1},z_{t-1}) \Big) \cdots \\ &  \hphantom{=}   \delta_{\max\{z,c(x,u_0)\}}(\mathrm{d}z_1) \; Q(\mathrm{d}x_1|x,u_{0}) \; \mu_{0}(\mathrm{d}u_{0}|x,z),\\
\end{aligned}\end{equation}
where there are $t$ copies of $C \times S \times \mathcal{Z}$ in \eqref{my1111}.
The term in parenthesis is $T_{\mu_{t-1}}(v_0^s)(x_{t-1},z_{t-1})$, and thus,
\begin{equation}\label{my14}\begin{aligned}
     J_{t,s}^\pi(x,z)   & = \textstyle \int_{C} \int_{S} \int_ {\mathcal{Z}} \cdots \Big( T_{\mu_{t-1}}(v_0^s)(x_{t-1},z_{t-1}) \Big) \cdots \\ & \hphantom{==} \delta_{\max\{z,c(x,u_0)\}}(\mathrm{d}z_1) \; Q(\mathrm{d}x_1|x,u_{0}) \; \mu_{0}(\mathrm{d}u_{0}|x,z)
\end{aligned}\end{equation}
with $t-1$ copies of $C \times S \times \mathcal{Z}$ in \eqref{my14}. By writing more of the integrals explicitly and using Definition \ref{DPoperator}, we find that
\begin{equation}\label{my15}\begin{aligned}
     & J_{t,s}^\pi(x,z) = \\
     &  \textstyle \int_{C} \int_{S} \int_ {\mathcal{Z}} \cdots \int_{C} \int_{S} \int_ {\mathcal{Z}} T_{\mu_{t-2}}\big( T_{\mu_{t-1}}(v_0^s)\big)(x_{t-2},z_{t-2}) \\ & \hphantom{=} \delta_{\max\{z_{t-3},c(x_{t-3},u_{t-3})\}}(\mathrm{d}z_{t-2}) \; Q(\mathrm{d}x_{t-2}|x_{t-3},u_{t-3})\\ & \hphantom{=} \mu_{t-3}(\mathrm{d}u_{t-3}|x_{t-3},z_{t-3}) \cdots \\ & \hphantom{=} \delta_{\max\{z,c(x,u_0)\}}(\mathrm{d}z_1) \; Q(\mathrm{d}x_1|x,u_{0}) \; \mu_{0}(\mathrm{d}u_{0}|x,z)
\end{aligned}\end{equation}
with $t-2$ copies of $C \times S \times \mathcal{Z}$ in \eqref{my15}. By repeating this process until the integral has one copy of $C \times S \times \mathcal{Z}$, we conclude that
\begin{equation}\begin{aligned}
    J_{t,s}^\pi(x,z)   & = \textstyle \int_{C} \int_{S} \int_ {\mathcal{Z}} T_{\mu_{1}}\big( \cdots \big( T_{\mu_{t-1}}(v_0^s)\big) \cdots \big)(x_{1},z_{1})  \\ & \hphantom{==} \delta_{\max\{z,c(x,u_0)\}}(\mathrm{d}z_1) \; Q(\mathrm{d}x_1|x,u_{0}) \; \mu_{0}(\mathrm{d}u_{0}|x,z)\\
   & = T_{\mu_{0}}( T_{\mu_{1}}( \cdots ( T_{\mu_{t-1}}(v_0^s) )\cdots ) )(x,z),
\end{aligned}\end{equation}
where we use Definition \ref{DPoperator} in the last line. If $t \in \mathbb{N}_0$ and $\pi = (\mu, \mu, \dots) \in \Pi$, then the previous results give
\begin{equation}\label{my1717}
    J_{t+1,s}^\pi(x,z) = T_{\mu}( T_{\mu} ( \cdots ( T_{\mu}(v_0^s) ) \cdots ))(x,z),
\end{equation}
where the operator $T_\mu$ is applied $t+1$ times. Using \eqref{my1717} and Definition \ref{DPoperator}, we have $J_{t+1,s}^\pi(x,z) = T_\mu(J_{t,s}^\pi)(x,z)$. 
\end{proof}
\section*{Acknowledgments}
The authors gratefully acknowledge Kevin M. Smith and Huizhen Janey Yu for discussions.
\bibliographystyle{IEEEtran}
\bibliography{references_new}

\end{document}